\documentclass[pra,onecolumn,showpacs,floatfix]{revtex4}
\usepackage{graphicx}
\begin{document}

\title{Dimensional reduction in Bose-Einstein condensed clouds of atoms confined in tight potentials of
any geometry and any interaction strength}
\author{P. Sandin$^1$, M. \"Ogren$^1$, M. Gulliksson$^1$} 
\affiliation{$^1$School of Science and Technology, \"Orebro University, 70182 \"Orebro, Sweden} 
\author{J. Smyrnakis$^2$, M. Magiropoulos$^2$, and G. M. Kavoulakis$^2$} 
\affiliation{$^2$Technological Education Institute of Crete, P.O. Box 1939, GR-71004, Heraklion, Greece}

\date{\today}

\begin{abstract}

Motivated by numerous experiments on Bose-Einstein condensed atoms which have been performed in tight 
trapping potentials of various geometries (elongated and/or toroidal/annular), we develop a general 
method which allows us to reduce the corresponding three-dimensional Gross-Pitaevskii equation for the 
order parameter into an effectively one-dimensional equation, taking into account the interactions 
(i.e., treating the width of the transverse profile variationally) and the curvature of the trapping 
potential. As an application of our model we consider atoms which rotate in a toroidal trapping potential. 
We evaluate the state of lowest energy for a fixed value of the angular momentum within various 
approximations of the effectively one-dimensional model and compare our results with the full solution 
of the three-dimensional problem, thus getting evidence for the accuracy of our model.

\end{abstract}
 \maketitle

\section{Introduction}

One of the many interesting and novel features of cold-atomic systems is the presence of a trapping 
potential. Compared to the homogeneous systems, these gases are finite, with a discrete energy spectrum. 
It is interesting that experimentalists can create trapping potentials which are both quasi-one- and 
quasi-two-dimensional, while more recently they have built annular and toroidal potentials. Clearly the 
reduced effective dimensionality of these systems, as well as the nontrivial topology of annular/toroidal 
traps introduce novel effects. 

Quasi-one-dimensional traps have been used in cold atomic systems ever since the first pioneering 
experiments and have given rise to very interesting effects. Already almost 20 years ago the propagation 
of sound waves in an elongated Bose-Einstein condensate was observed and studied experimentally \cite{Kett}. 
Soon after solitary waves were observed in an elongated trap \cite{sol1} and in a more oblate one \cite{sol2}. 
More recently toroidal and annular traps have been built, and even persistent currents have been observed, 
see, e.g., \cite{Kurn, Olson, Phillips1, Foot, GK, Moulder, Ryu, Zoran, hysteresis}. Remarkably, it has also 
become possible to manipulate the shape of the trapping potentials to a very high degree, see, e.g., 
Ref.\,\cite{Bosh}.

Under typical conditions the non-linear Gross-Pitaevskii equation provides a very accurate description of 
these systems. In principle this equation may be solved numerically with use of various techniques, but 
still the full three-dimensional problem may become rather challenging. When the trapping potential is very 
tight along two, or one dimension (i.e., the quantum of energy along the corresponding direction is much 
larger than any other energy scale of the problem) the motion of the atoms is quasi-one, or quasi-two-dimensional, 
respectively. In this case one may make an ansatz for the order parameter, assuming a decoupling of the degrees 
of freedom along the direction of tight confinement, which are frozen. Integrating over these degrees of 
freedom one may derive an effectively one-, or two-dimensional equation for the order parameter. Such efforts 
have been made by numerous authors. Given the large numbers of these references we refer just to some 
representative ones, \cite{KP, JKP, LP, Sal, Car}.

The benefit from such effective theories is two-fold. First of all, the corresponding numerical problem is
easier to solve. Secondly, especially in the case of quasi-one-dimensional motion, one has a handy equation
to work with, which may provide insight into the problem. In addition, one may derive simple expressions 
for various observables, thus making contact with the well-known corresponding expressions of the homogeneous 
systems (i.e., the one where there is no trapping potential).

Motivated by the numerous experiments which have been performed in such traps, as well as by the arguments 
presented above, we consider quite generally a tight, quasi-one-dimensional potential of any geometry and
develop an effectively one-dimensional theory. This is the first main result of the present study. Then, 
we apply our model to the problem of atoms which rotate in a toroidal trapping potential. We solve numerically 
the one-dimensional problem under various approximations, i.e., the purely one-dimensional problem, the 
quasi-one-dimensional problem with a fixed transverse width, and the quasi-one-dimensional problem with the 
transverse width treated variationally. In addition we solve the full three-dimensional problem numerically, 
thus making a quantitative comparison of the various models with the exact solution, which is our second main
result.  

In what follows below we present in Sec.\,II our quasi-one-dimensional model, while in Sec.\,III we give 
approximate and limiting expressions that result from it. In Sec.\,IV we show the corresponding numerical 
solutions that result within each approximation, as well as the results of the full three-dimensional problem 
and compare them. Finally in Sec.\,V we summarize our results and give an overview.

\section{Derivation of the effective 1D model}

The starting point is the (time-independent) Gross-Pitaevskii equation in three dimensions, 
\begin{equation}
-\frac{1}{2} \nabla^2 \Phi + V \Phi + N U_0 |\Phi|^2 \Phi - \mu \Phi = 0, 
\label{NLSE}
\end{equation}
where $\Phi$ is the order parameter, $N$ is the atom number, $V$ is the trapping potential, $U_0$ is the matrix 
element for atom-atom collisions, and $\mu$ is the chemical potential (the mass of the atoms $M$, as well as $\hbar$ 
are set equal to unity, while $\Phi$ is normalized to unity).

Let us now suppose that we want to solve the above three-dimensional Gross-Pitaevskii equation in a narrow tubular 
neighbourhood of a planar curve $\underline{x}_0(s)$. We use the following parametrization, 
\begin{equation}
\underline{x}(s,\rho,\phi ) = \underline{x}_0(s)+ \rho \sin\phi \, \hat{n}(s )+
\rho \cos \phi \, \hat{n}(s) \times \hat{T}(s).
\label{param}
\end{equation}
The schematic plot of Fig.\,1 shows how $\rho$, $\phi$, and $s$ are defined. In Eq.\,(\ref{param}) ${\hat T}(s)$ 
is the tangent and ${\hat n}(s)$ is the outer normal to the curve $\underline{x}_0(s)$. It is convenient to set 
$s$ equal to the arclength from a fixed origin of the curve. If we do this, ${d\underline{x}_0(s )}/{ds}=\hat{T}(s)$, 
${d\hat{T}(s)}/{ds}=-k(s) \hat{n}(s)$ and ${d\hat{n}(s)}/{ds} = k(s )\hat{T}(s )$, where $k(s)$ is the curvature of 
the curve. In the above coordinates the metric $dw^2$ becomes
\begin{eqnarray}
dw^2&=&d\underline{x}\cdot d\underline{x}= \left| \frac {d \underline{x}} {d s} \right|^2 d s^2
+ \left| \frac {d \underline{x}} {d \rho} \right|^2 d \rho^2
+ \left| \frac {d \underline{x}} {d \phi} \right|^2 d \phi^2
\nonumber \\
&=& [1+k(s)\rho \sin\phi]^2ds^2+d\rho^2+\rho^2 d\phi^2.
\label{metric}
\end{eqnarray}
Equation (\ref{NLSE}) thus becomes,
\begin{eqnarray}
&-&\frac{1}{2}\frac{1}{1+k(s )\rho \sin\phi }\Big{[} \frac{\partial}{\partial
s }\left( \frac{1}{1+k(s )\rho \sin\phi }
\frac{\partial \Phi}{\partial s }\right)
+\frac 1 {\rho} \frac{\partial }{\partial \rho }\left( \rho [1+k(s)\rho \sin\phi]
\frac{\partial \Phi }{\partial \rho }\right) \nonumber \\
&+& \frac 1 {\rho} \frac{\partial} {\partial \phi} \left( [1+k(s)\rho \sin\phi] \frac{\partial \Phi }{\rho \partial \phi }
\right) \Big{]}
+ V(\rho) \Phi + N U_0 |\Phi|^2 \Phi - \mu \Phi = 0,
\label{3dNLS}
\end{eqnarray}
where the normalization for $\Phi$ has the form 
\begin{eqnarray}
  \int |\Phi|^2 [1+k(s)\rho \sin\phi] \, \rho d \rho d s d \phi = 1. 
\end{eqnarray}
The trapping potential $V$ in the problem that we have in mind is assumed to act transversely to the curve 
and is a function of $\rho$ only, while it is taken to be harmonic, $V(\rho)=\omega^2\rho^2/2$.  

\begin{figure}
\includegraphics[width=6cm,height=5cm,angle=-0]{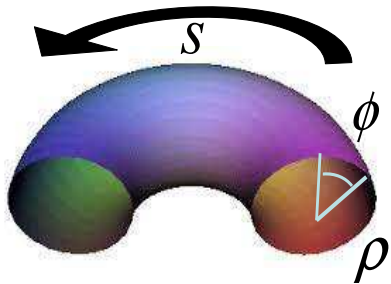}
\vskip1pc
\caption{(Colour online) Schematic picture, which shows the coordinates $\rho$, $\phi$, and $s$.}
\end{figure}

Setting $\Psi = \sqrt{h(\rho, s,\phi)} \Phi$, where $h \equiv 1+k(s )\rho \sin\phi$ we get for $\Psi$
\begin{eqnarray}
  - \frac 1 2 \frac {\partial} {\partial s} \left( \frac {{\Psi}_{s}} {h^2} \right)  
  - \frac 1 2 \frac 1 {\rho} \frac {\partial} {\partial \rho} (\rho {\Psi}_{\rho}) 
  - \frac 1 2 \frac 1 {\rho^2} {\Psi}_{\phi \phi} 
  - \frac 1 8 \frac {k^2(s)} {h^2} {\Psi} 
  - \frac 5 8 \frac {h^2_{s}} {h^4} {\Psi}
  + \frac 1 4 \frac {h_{s s}} {h^3} {\Psi} 
  + V(\rho) {\Psi}  
  + N U_0 h |{\Psi}|^2 {\Psi} 
  - \mu \Psi = 0,
\label{neq}
\end{eqnarray}
where the low index on the right denotes differentiation with respect to the corresponding variable. The 
Hamiltonian that corresponds to Eq.\,(\ref{neq}) is
\begin{eqnarray}
H = \int 
 \left[
    \frac 1 2 \frac {|{\Psi}_{s}|^2} {h^2}  
  + \frac 1 2 |{\Psi}_{\rho}|^2
  + \frac 1 2 \frac 1 {\rho^2} |{\Psi}_{\phi}|^2 
  - \frac 1 8 \frac {k^2(s)} {h^2} |{\Psi}|^2 
  - \frac 5 8 \frac {h^2_{s}} {h^4} |{\Psi}|^2
  + \frac 1 4 \frac {h_{s s}} {h^3} |{\Psi}|^2 
  \right.
\nonumber \\
\left. 
+ V(\rho) |{\Psi}|^2 + \frac 1 2 \frac 1 h N U_0 |{\Psi}|^4 - \mu |{\Psi}|^2 \right] 
\rho d \rho ds d\phi.
\label{hamil}
\end{eqnarray}

Up to now the calculation we have performed is exact. In what follows below we will make the following ansatz
\begin{equation}
{\Psi}(s, \rho, \phi) = \psi_{\rm tr}(\rho, a_{\perp}(s)) \cdot \psi(s),
\label{ansatz}
\end{equation}
where 
\begin{equation}
\psi_{\rm tr}(\rho, a_{\perp}(s))=\frac{1}{\sqrt{\pi} a_{\perp}(s)} e^{-\rho^2/2 a_{\perp}^2(s)}
\label{ass2}
\end{equation}
is the normalized, rotationally symmetric, ground state of $V(\rho)$, with a spatially-dependent width 
$a_{\perp}(s)$, which is attributed to the (density-dependent) nonlinear term. The above ansatz assumes that 
the order parameter is independent of the azimuthal variable $\phi$. In addition, within this ansatz the 
functional form of the transverse profile of the cloud is Gaussian, however the cloud is allowed to expand 
radially, rather than having a fixed width, which would be the oscillator length $a_0 = 1/\sqrt \omega$.

In Ref.\,\cite{JKP} a closely-related problem has been solved, namely the one where there is no curvature, 
$k(s) = 0$, making the same ansatz as that of Eq.\,(\ref{ansatz}). In the limit of weak interactions the actual
transverse profile is Gaussian, while in the Thomas-Fermi limit the profile is an inverse parabola. Here, since 
we want our model to be applicable for any interaction strength we have chosen to work with a Gaussian for all 
interaction strengths, however its width $a_{\perp}$ is treated variationally and it is allowed to increase when 
the interaction is sufficiently strong. Therefore, viewed as a variational approach, the present model is expected 
to work rather well, as we confirm in Secs.\,III and IV. 

The authors of Ref.\,\cite{Car} have attacked the same problem as the present one, including the effect of the 
curvature. However, in Ref.\,\cite{Car} the transverse profile has a fixed width, i.e., $a_{\perp}$ has been set 
equal to the oscillator length $a_0$. Thus, for zero, or weak interactions the two approaches coincide. However, 
our model is more general and it works better when the interaction becomes comparable to, or larger than, the 
oscillator quantum of energy $\omega$, as seen in the evaluation of the speed of sound in Sec.\,III and in the
explicit problem considered in Sec.\,IV.

Performing the integration in Eq.\,(\ref{hamil}) over the variable $\phi$, we find that
\begin{eqnarray}
H = \int 
 \left[
    \frac 1 2 \frac 1 {[1 - k^2(s) \rho^2]^{3/2}} |\Psi_{s}|^2  
  + \frac 1 2 |{\Psi}_{\rho}|^2
  - \frac 1 8 \frac {k^2(s)} {[1 - k^2(s) \rho^2]^{3/2}} |{\Psi}|^2 
  - \frac 5 8 \frac {A(\rho,s)} {2 \pi} |{\Psi}|^2
  + \frac 1 4 \frac {B(\rho,s)} {2 \pi} |{\Psi}|^2 
  \right.
\nonumber \\
\left. 
+ \frac 1 2 \frac {N U_0} {\sqrt{1 - k^2(s) \rho^2}} |{\Psi}|^4 + V(\rho) |{\Psi}|^2 - \mu |{\Psi}|^2 \right] 
2 \pi \rho d \rho ds,
\label{hamill}
\end{eqnarray}
where \cite{integrals}
\begin{eqnarray}
A = - [k_s(s) \rho]^2 \frac {3 \pi k(s) \rho} {{[1 - k^2(s) \rho^2]}^{3/2}},
\end{eqnarray}
and 
\begin{eqnarray}
B = k_{ss}(s) \pi \rho \frac {1 + 4 k^2(s) \rho^2} {[1 - k^2(s) \rho^2]^{7/2}}.
\end{eqnarray} 
Equation (\ref{hamill}) leads to the $\phi$-independent equation
\begin{eqnarray}
  - \frac 1 2 \frac {\partial} {\partial s} \left[ \frac 1 {[1 - k^2(s) \rho^2]^{3/2}} \Psi_{s} \right]
  - \frac 1 2 \frac 1 {\rho} \frac {\partial} {\partial \rho} (\rho \Psi_{\rho}) 
  - \frac 1 8 \frac {k^2(s)} {[1 - k^2(s) \rho^2]^{3/2}} \Psi
  - \frac 5 8 \frac {A(\rho, s)} {2 \pi} \Psi
  + \frac 1 4 \frac {B(\rho, s)} {2 \pi} \Psi +
  \nonumber \\
  + \frac {N U_0} {\sqrt{1 - k^2(s) \rho^2}} |\Psi|^2 \Psi + V(\rho) \Psi - \mu \Psi = 0.
\end{eqnarray}
Up to now we have only made use of the independence of the order parameter from $\phi$ and not of the factorization
in the form of a longitudinal and a transverse function. 

Before we proceed, it is instructive to mention that there are three relevant length scales in the present problem, 
namely (i) the oscillator length $a_0$ -- that sets the scale for the transverse width of the cloud $a_{\perp}$, 
(ii) the coherence (or healing) length $\xi$, that is defined as $1/(2 \xi^2) = n_0 U_0$, where $n_0$ is the 
three-dimensional density of the homogeneous system and it sets the scale of the solitary-wave, travelling wave 
solutions when the potential is sufficiently tight, and (iii) the length scale associated with the curvature, 
$k^{-1}$, that determines the scale over which the trapping potential ``bends". 

We now use the factorization made in the ansatz of Eq.\,(\ref{ansatz}), as well as Eq.\,(\ref{ass2}). Furthermore, 
we assume that $a_{\perp} \ll |k^{-1}|$ and $a_{\perp} \ll \xi$. These conditions imply that $|k| a_{\perp} \ll 1$ 
and $({a_{\perp}})_{s} \ll 1$. Thus, we find the effectively one-dimensional Hamiltonian
\begin{eqnarray}
 H_{1d} = \int \left[ \frac 1 2 |\psi_s|^2 + \frac 1 {2 a_{\perp}^2} |\psi|^2 
 - \frac 1 8 k^2(s) |\psi|^2 
 + \frac 1 2 \frac {N U_0} {2 \pi a_{\perp}^2} |\psi|^4 + \frac 1 2 \omega^2 a_{\perp}^2 |\psi|^2 - \mu |\psi|^2 \right] \, 
 d s,
\end{eqnarray}
where $\int |\psi|^2 \, d s = 1$. Demanding that variations in $\psi^*$ and in $a_{\perp}^2$ in the above functional 
vanish we find that
\begin{eqnarray}
- \frac {1} {2} \psi_{ss}
 + \frac 1 {2 a_{\perp}^2} \psi - \frac 1 8 k^2(s) \psi + \frac {N U_0} {2 \pi a_{\perp}^2} |\psi|^2 \psi
 + \frac 1 2 \omega^2 a_{\perp}^2 \psi - \mu \psi = 0,
\label{equ1}
\end{eqnarray}
and
\begin{eqnarray}
  a_{\perp}^4 = \frac 1 {\omega^2} \left[1 + \frac {N U_0 |\psi|^2} {2 \pi} \right] = 
  a_0^4 \left[1 + \frac {N U_0 |\psi|^2} {2 \pi} \right].
\label{variational}
\end{eqnarray}
The coefficient of the nonlinear term in Eq.\,(\ref{equ1}) may be identified as the integral of $|\psi_{\rm tr}(\rho)|^4$ 
over the cross section of the cloud, $U_0 \int |\psi_{\rm tr}(\rho)|^4 \rho d \rho d \phi = U_0/(2 \pi a_{\perp}^2)$. The 
dimensionless quantity $N U_0 |\psi|^2/(2 \pi)$ in Eq.\,(\ref{variational}) is equal to the ratio between the 
interaction energy and the quantum of energy $\omega$. If $a_{\rm sc}$ is the scattering length for elastic atom-atom 
collisions, then $U_0 = 4 \pi a_{\rm sc}$, and therefore $N U_0 |\psi|^2/(2 \pi) = 2 \sigma a_{\rm sc}$, where $\sigma 
\equiv N |\psi|^2$ is the density per unit length. 

Combining Eqs.\,(\ref{equ1}) and (\ref{variational}) we get that
\begin{eqnarray}
 - \frac {1} {2} \psi_{ss}
 + \left[ \frac {\omega} {2 \sqrt{1 + N U_0 |\psi|^2/(2 \pi)}} - \frac 1 8 k^2(s) 
 + \frac 1 2 \omega \sqrt{1 + N U_0 |\psi|^2/(2 \pi)} 
 + \frac {N U_0 \omega} {2 \pi \sqrt{1 + N U_0 |\psi|^2/(2 \pi)}} |\psi|^2 - \mu \right] \psi = 0.
 \label{fineq}
\end{eqnarray}
The above equation is the first main result of the present study. The initial problem, which has three spatial 
dimensions, has been reduced to a problem of one dimension. We stress the generality of the above equation, 
where for some given $k(s)$ -- i.e., a trapping geometry of any shape -- and also for any interaction strength, 
these equations are applicable (under the assumptions that we mentioned above). In addition, it is interesting that 
one may identify separately the effect of the curvature, of the transverse confinement, and of the interaction. 

\section{Approximations and applications of the effective 1D model}

It is instructive to examine some limiting cases of our model, which also allows us to get some insight. First of
all, Eq.\,(\ref{variational}) is an algebraic equation for the transverse width of the cloud, $a_{\perp}$. In the 
absence of interactions $a_{\perp}$ is equal to the oscillator length $a_0$, as expected. When the system is 
in the Thomas-Fermi regime, i.e., the interaction energy is much larger than $\omega$, then $a_{\perp}/a_0 \approx 
(2 \sigma a_{\rm sc})^{1/4}$. In this limit the width of the cloud thus increases due to the interactions \cite{KP, 
JKP}, since $\sigma a_{\rm sc} \gg 1$. We should stress that in the Thomas-Fermi limit the coherence length may 
become smaller than the oscillator length and our model approaches its limits of validity \cite{JKP}.

Equation (\ref{fineq}) has various interesting limits. For weak interactions \cite{Car}, 
\begin{eqnarray}
 - \frac {1} {2} \psi_{ss}
 + \left[- \frac 1 8 k^2(s) 
 +\frac {N U_0} {2 \pi a_0^2} |\psi|^2 + \omega - \mu \right] \psi = 0.
\label{ap1}
\end{eqnarray}
In the case of a ring-like potential, where $k(s)$ is constant and equal to $1/R$, with $R$ being the radius of the 
ring, the effect of the curvature is trivial, since the above equation essentially coincides with the purely 
one-dimensional equation.

In the opposite limit of ``strong" interactions, i.e., in the Thomas-Fermi limit the nonlinear term in the effective 
equation, Eq.\,(\ref{fineq}), has a different functional form, which is $|\psi| \psi$, instead of the usual one, 
$|\psi|^2 \psi$ \cite{JKP},
\begin{eqnarray}
 - \frac {1} {2} \psi_{ss}
 + \left[ - \frac 1 8 k^2(s) 
 + \frac 3 2 \omega \sqrt{\frac {N U_0} {2 \pi}} |\psi| - \mu \right] \psi = 0.
 \label{fineqq}
\end{eqnarray}
Setting $k(s) = 0$ and after linearisation one gets the following Bogoliubov spectrum from Eq.\,(\ref{fineqq})
\begin{eqnarray}
  E(p) = \sqrt{ \frac 1 4 p^4 + \frac 3 4 \left( \frac {\sigma_0 U_0} {2 \pi} \right)^{1/2} \omega \, p^2},
\end{eqnarray}
for the energy $E$, where $p$ is the momentum and $\sigma_0$ is the background, one-dimensional density, i.e., the
density at $|s| \to \infty$. For $p \to 0$ we get a speed of sound which is
\begin{eqnarray}
  c^2 = \frac 3 4 \omega \sqrt{\frac {\sigma_0 U_0} {2 \pi}} = \frac {3 \sqrt 2} 4 \omega \sqrt{\sigma_0 a_{\rm sc}}.
\end{eqnarray}
The above expression should be compared with the exact result \cite{JKP} (i.e., the one with the inverse parabola 
for the transverse profile, instead of the Gaussian that is assumed here),
\begin{eqnarray}
  c_{\rm ex}^2 = \omega \sqrt{\frac {\sigma_0 U_0} {4 \pi}} = \omega \sqrt{\sigma_0 a_{\rm sc}},
\end{eqnarray}
and thus $(c/c_{\rm ex})^2 = (3 \sqrt 2/ 4) \approx 1.06$. As expected, the speed of sound in the approximate scheme 
is higher, however the difference (roughly $3\%$ for the velocities) is rather small. On the other hand, if one uses 
Eq.\,(\ref{ap1}) instead (thus ignoring the effect of the interaction on the width of the transverse profile), then 
$(c/c_{\rm ex})^2 = 2 \sqrt{\sigma_0 a_{\rm sc}}$, which is $\gg 1$ in the Thomas-Fermi limit, and thus this model 
fails.

\section{Numerical results}

\subsection{Bosonic atoms confined in a tight toroidal trap}

Having established the general framework which allows us to reduce the three-dimensional problem into an effectively 
one-dimensional equation, we now apply our method to the problem of a toroidal trapping potential, and in particular 
to the problem where the atoms are given some angular momentum. In the case of an axially-symmetric (toroidal) geometry 
$k(s) = 1/R$. To make a comparison between the various models, first of all, we solve the purely one-dimensional 
problem, 
\begin{eqnarray}
-\frac{1}{2 R^2} \frac{d^2 \psi}{d \theta^2}
+ \left[ \frac{N U_0}{2\pi a_0^2} |\psi|^2 + \omega - \mu \right] \psi
+ i \Omega \frac {d \psi} {d \theta}= 0.
\label{weaktightNLSSS}
\end{eqnarray}
Here $\Omega$ is a Lagrange multiplier that takes care of the angular momentum and $\theta = s/R$. This equation is 
solved under the constraints of a fixed atom number and a fixed value of the angular momentum; for more details see 
Ref.\,\cite{SOG}.
   
We also solve the same problem (i.e., under the same constraints of a fixed atom number and a fixed value of the 
angular momentum, with $k(s) = 1/R$) using Eq.\,(\ref{equ1}), setting the transverse width $a_{\perp}$ equal to 
$a_0$. As we mentioned earlier, this approximation for the transverse width being constant and equal to $a_0$ has 
been made in Ref.\,\cite{Car}. Finally, we solve the same problem using Eqs.\,(\ref{equ1}) and (\ref{variational}), 
i.e., we solve Eq.\,(\ref{fineq}), thus treating the transverse width of the cloud $a_{\perp}$ variationally. 

The plots in Fig.\,2 show the dispersion relation within (i) the strictly one-dimensional model, (ii) the 
effective one-dimensional model with a fixed transverse width (corresponding to the calculation of Ref.\,\cite{Car}), 
(iii) the effective one-dimensional model with a variable transverse width, and finally (iv) the energy of the 
full three-dimensional solutions, for three values of $\omega$ and $U_0$ (as explained in detail in the following 
section). The energy between the first and the second differ by the constant factor $\Delta E = -k^2/8 = -1/8$ (for 
$R = 1$ that we have assumed) due to the constant curvature term in the energy. As the ratio $n_0 U_0/\omega$ 
decreases the system approaches the one-dimensional limit. Indeed, as seen from these three figures, as this ratio 
decreases all the curves come closer to each other. More importantly, for the largest ratio of $n_0 U_0/\omega$, 
where there are substantial deviations from the one-dimensional limit, our model with a variable width provides an 
accurate description of the energy of the full, three-dimensional problem. 

The plots in Fig.\,3 show the density of the order parameter for two values of the angular momentum $\ell = 1/2$
and $\ell = 3/4$. In order for the effect of the interaction to become more pronounced, we set the ratio $n_0 U_0/
\omega = 5/2$. For this choice of parameters, $\xi/a_0 = 1/\sqrt 5 \approx 0.447$, $R/a_0 = \sqrt{20} \approx 4.472$, 
while $\xi/R = 1/10$. Within the purely one-dimensional model the solution with $\ell = 1/2$ corresponds to a ``dark" 
solitary wave and thus there is a node in the density, as seen in Fig.\,5. This feature is preserved within our 
quasi-one-dimensional model with a variable transverse width. In addition, for both values of $\ell$ we observe that 
the width of this solution is larger than the one of the purely one-dimensional model.

\begin{figure}
\includegraphics[width=8cm,height=8cm,angle=-0]{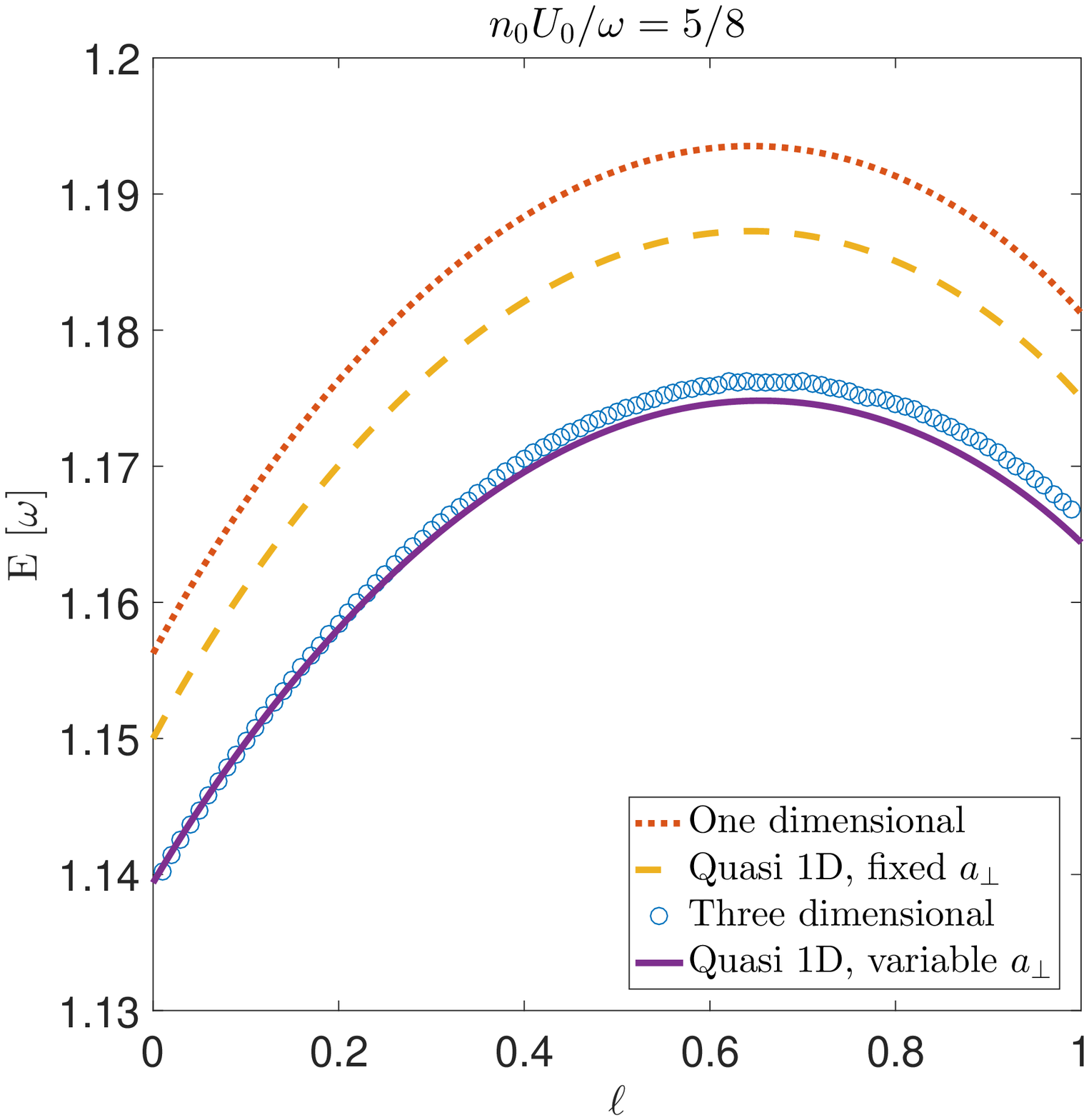}
\includegraphics[width=8cm,height=8cm,angle=-0]{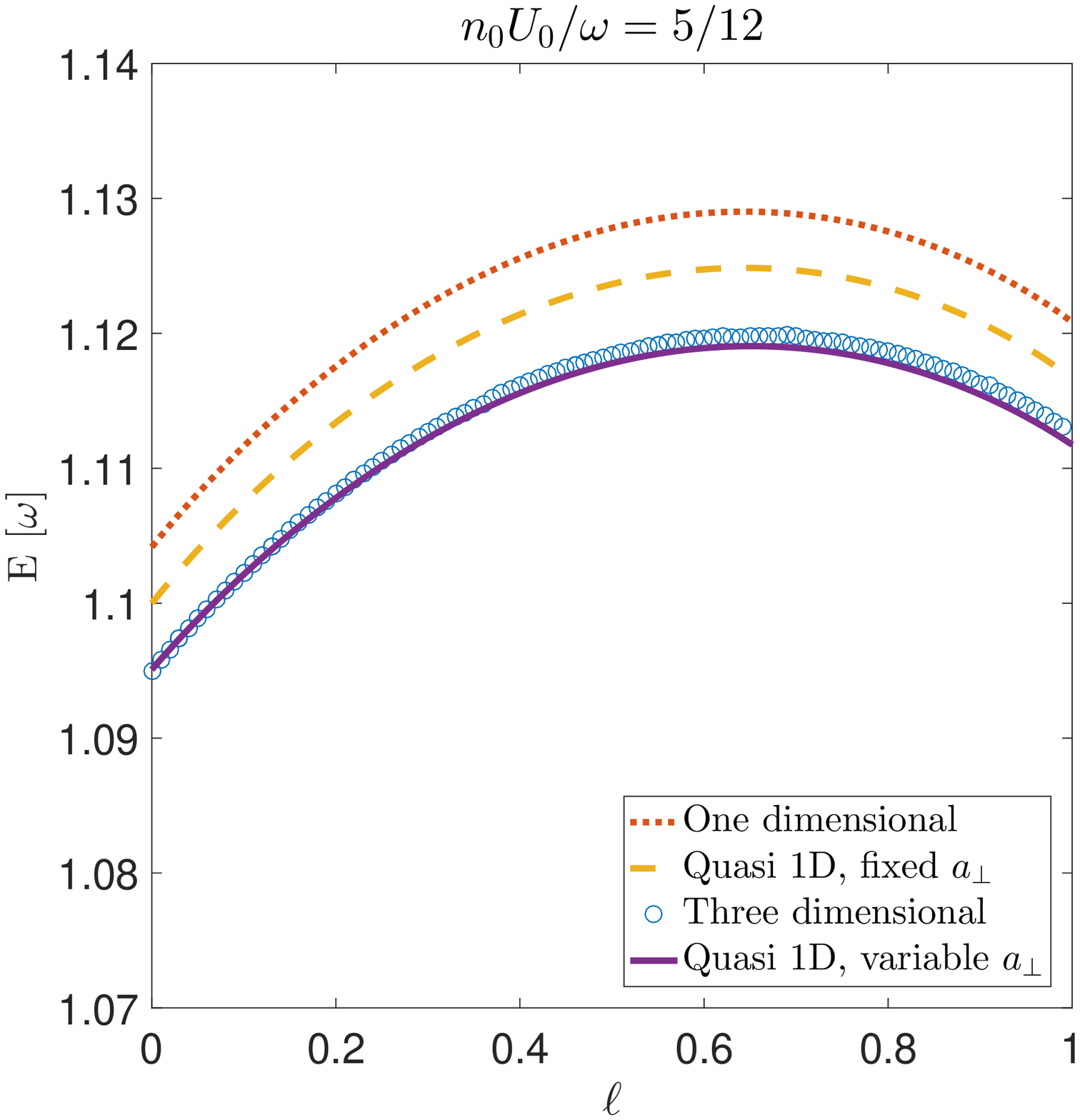}
\includegraphics[width=8cm,height=8cm,angle=-0]{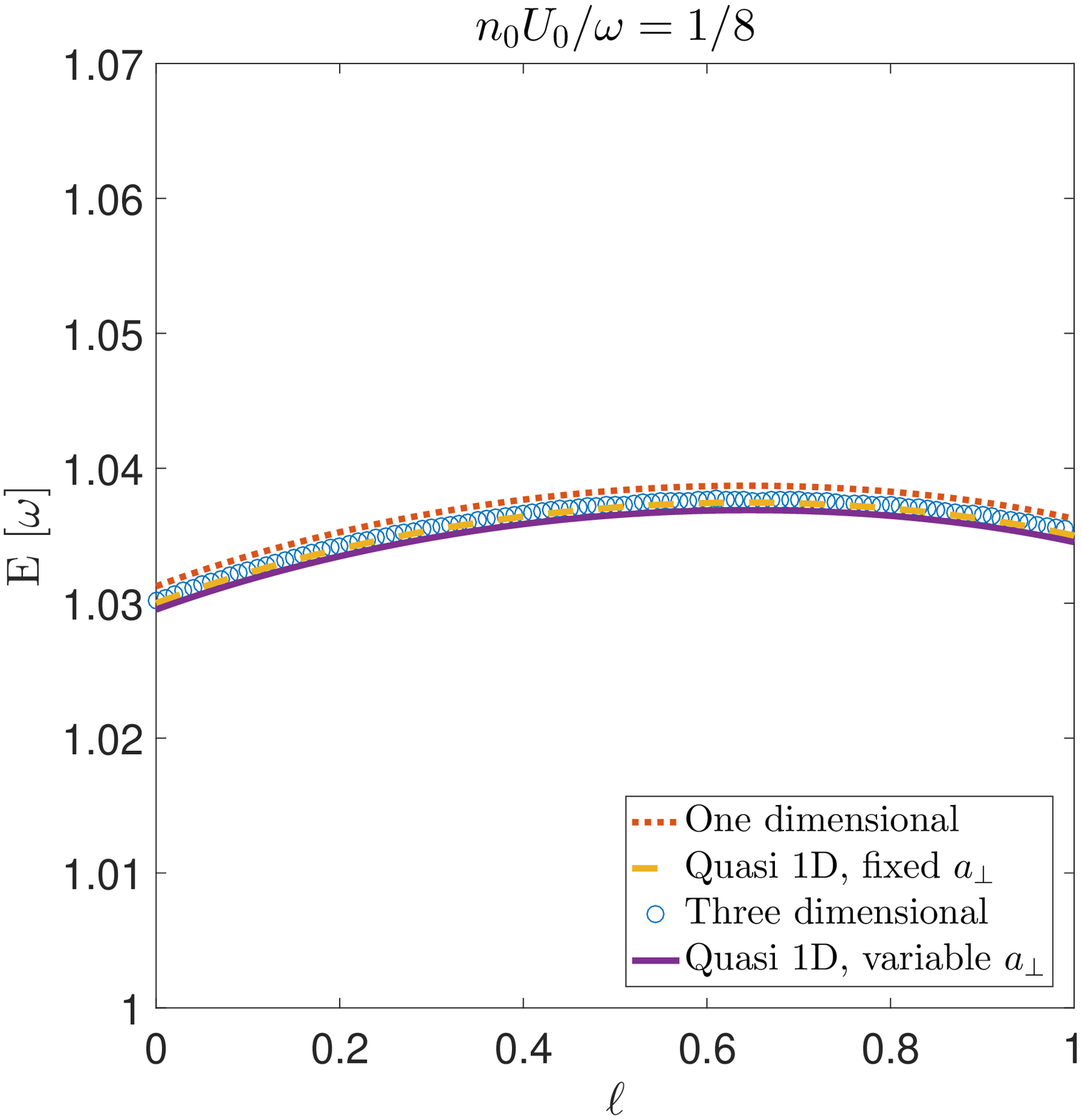}
\vskip1pc
\caption{(Colour online) The dispersion relation, i.e., the energy versus the angular momentum per particle $\ell$, 
(i) within the strictly one-dimensional model, (ii) within the effective one-dimensional model with a fixed transverse 
width (corresponding to the calculation of Ref.\,\cite{Car}), (iii) within the effective one-dimensional model with a 
variable transverse width, and finally (iv) within the full, three-dimensional problem. Here $R/\xi = 5$ in all
the plots, and also $\xi/a_0 = 2/\sqrt 5 \approx 0.894$, $R/a_0= \sqrt{20} \approx 4.472$ (upper left); $\xi/a_0 = 
1.095$, $R/a_0 = \sqrt{30} \approx 5.477$ (upper right); and $\xi/a_0 = 2$, $R/a_0 = 10$ (lower).}
\end{figure}

\begin{figure}
\includegraphics[width=8cm,height=8cm,angle=-0]{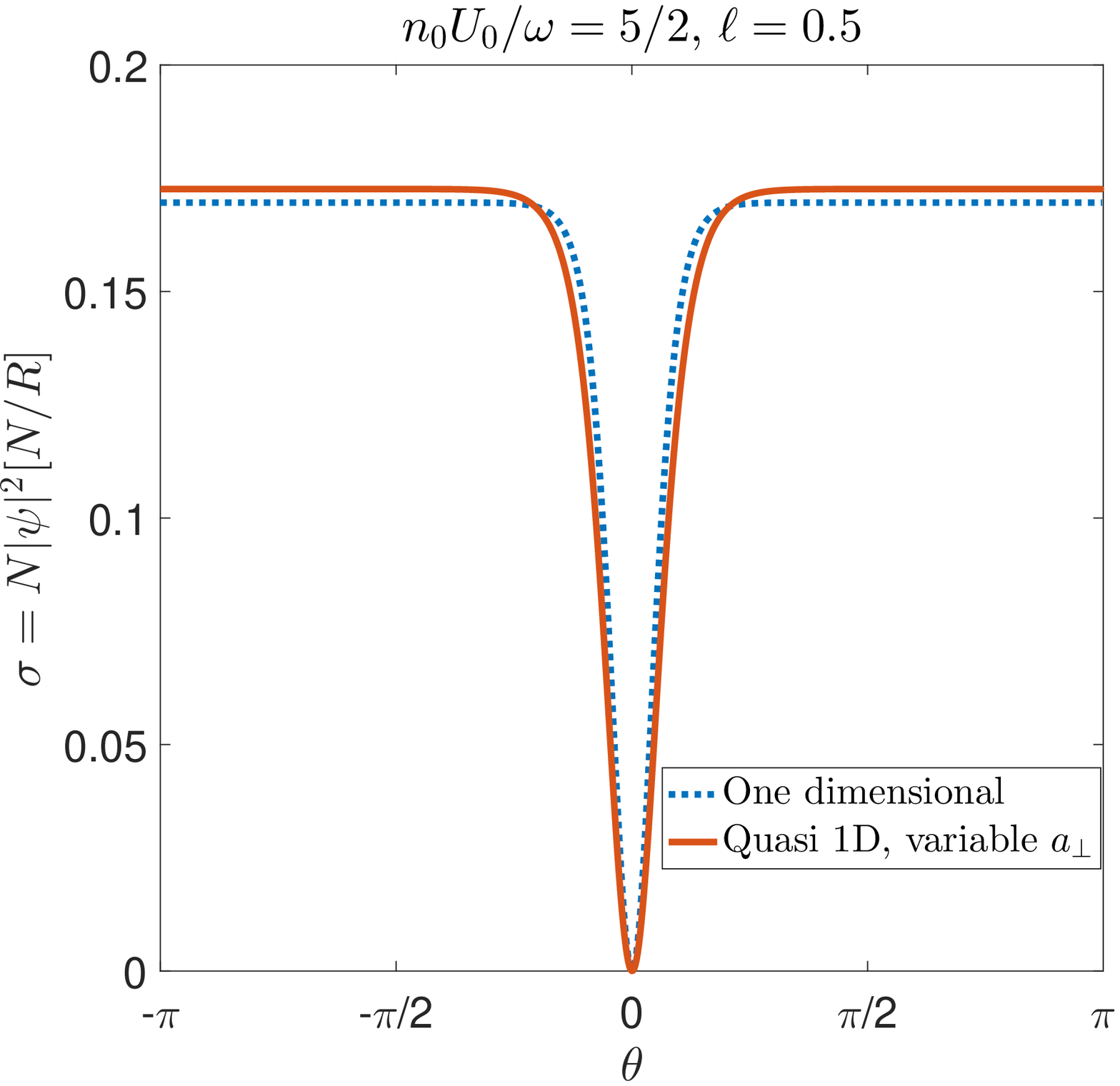}
\includegraphics[width=8cm,height=8cm,angle=-0]{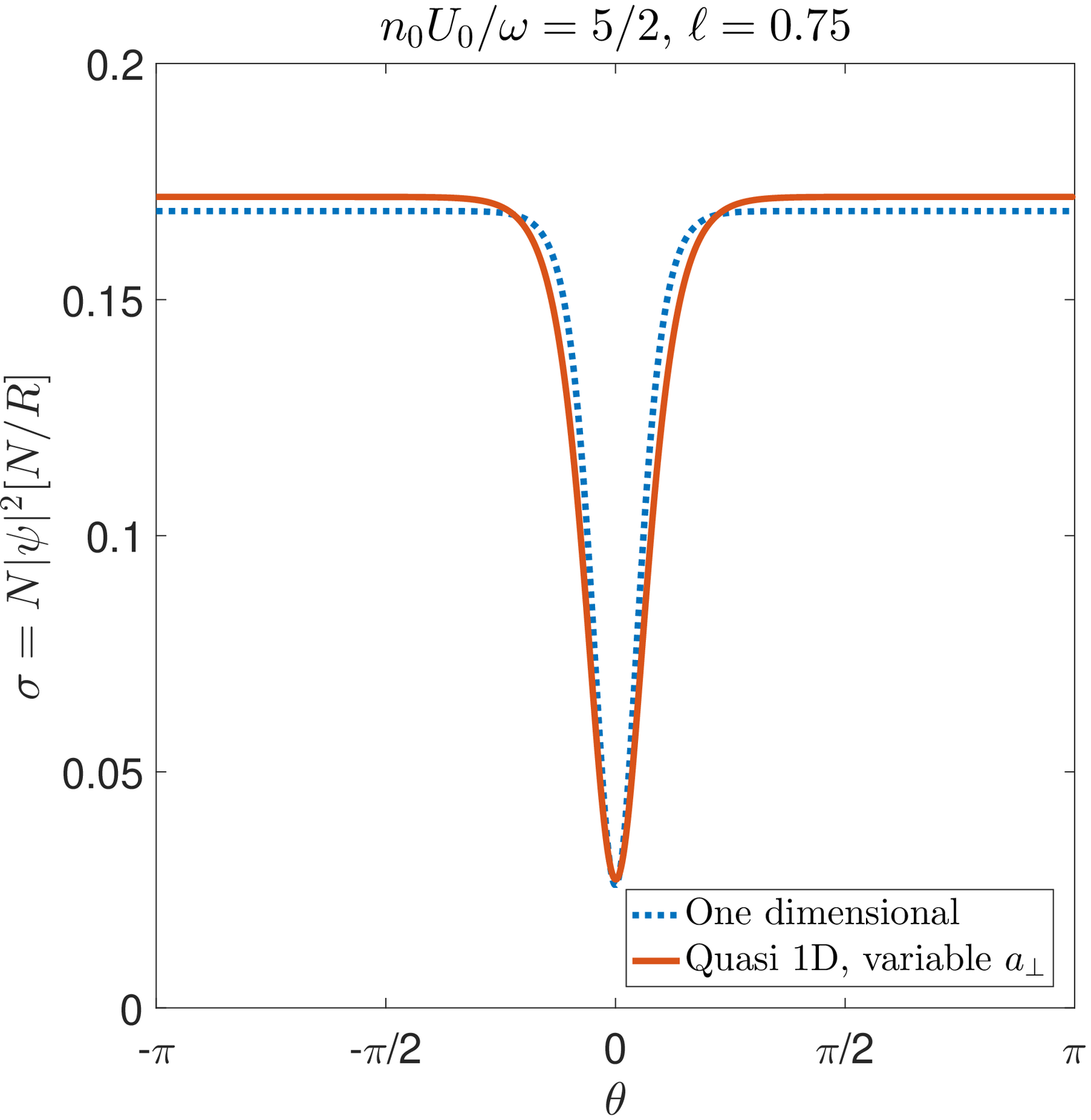}
\vskip1pc
\caption{(Colour online) The density per unit length $\sigma = N |\psi|^2$ of the order parameter $\psi$ as function 
of the angle $\theta$ within the purely one-dimensional model, and within the effective one-dimensional model with a 
variable transverse width for $\xi/a_0 = 1/\sqrt 5 \approx 0.447$, $R/a_0 = \sqrt{20} \approx 4.472$, and $\xi/R = 
1/10$, for $\ell = 1/2$ (left), and $\ell = 3/4$ (right).}
\end{figure}

\subsection{Full, three-dimensional problem and comparison with the effective one-dimensional models}

To make a comparison with an ``exact" problem, we also solve the full, three-dimensional problem. To do this, we
consider a trapping potential that has the form (in cylindrical coordinates)
\begin{equation}
V(\rho, \phi, z) = \frac 1 2 \omega^2 [(\rho-R)^2 + z^2]. 
\label{potential}
\end{equation}
We impose the same constraints as above, namely a fixed atom number and a fixed value of the angular momentum.

Since one of the main purposes of our study is to identify the effect of the deviations from purely one-dimensional
motion, first of all, we need to identify this limit. This limit is achieved when the interaction energy $n_0 U_0$
is much less that $\omega$. For weak interactions, $n_0 = N/(2 \pi R \pi a_0^2)$, and thus $n_0 U_0/\omega = N U_0/
(2 \pi^2 R)$. We stress that this ratio does not depend on $\omega$, since the density scales as $a_0^{-2} = \omega$. 
Finally, the corresponding ratio $\xi/a_0$ is $\sqrt{\pi^2 R/(N U_0)}$.

The other relevant ratio is that between $n_0 U_0$ and the kinetic energy $K$ for motion of the atoms along a ring 
with radius $R$, i.e., $K = 1/(2 R^2)$. The ratio $n_0 U_0/K$ is given by $N U_0 \omega R/\pi^2$. In addition, 
$\xi/R = \pi/\sqrt{N U_0 \omega R}$. For fixed $N$ and $R$, the limit of one-dimensional motion is the one where 
$U_0$ is ``small" (in the sense described above), with $n_0 U_0/K = N U_0 \omega R/\pi^2$ fixed, which implies that 
the product $U_0 \omega$ has to be kept fixed. 

From the arguments described above, we choose three sets of parameters. In all of them we set $n_0 U_0/K = 25$, or 
in terms of length scales $\xi/R = 1/5$. In the first set $n_0 U_0/\omega = 5/8 = 0.625$, or in terms of length 
scales $\xi/a_0 = 2/\sqrt 5 \approx 0.894$, while $R/a_0 = \sqrt{20} \approx 4.472$. In the second intermediate 
set the system is closer to one-dimensional motion. Here, $n_0 U_0/\omega = 5/12 \approx 0.417$, or $\xi/a_0 = 
1.095$, and $R/a_0 = \sqrt{30} \approx 5.477$. Finally, in the third set we are even closer to the one-dimensional 
limit, $n_0 U_0/\omega = 1/8 = 0.125$, or in terms of length scales $\xi/a_0 = 2$, while $R/a_0 = 10$. In the first 
two cases the coherence length is roughly equal with the oscillator length, while in the third it is twice as large. 

Especially in the first case the transverse degrees of freedom start to play a role. We stress that the condition 
for quasi-one-dimensional motion implies that the interaction energy does not exceed the oscillator quantum of energy 
in the transverse direction (an equivalent way of expressing this condition is that the coherence length, which sets 
the scale of the width of the solitary-wave profiles, is at least on the order of, or larger than the transverse width 
of the cloud, which is set by the oscillator length.) Still, even in the first case, where we have come closer to 
breaking the limits of validity of our model, we see in Fig.\,2 that our results are rather accurate.    

The coupled three-dimensional equations are solved using the built-in stationary solver in the commercial 
software COMSOL Multiphysics, version 5.2a. The domain is set up as a torus with major radius $R = 1$, and minor 
radius $4 a_0 = 4/\sqrt{\omega}$. This minor radius of the computational domain is deemed sufficiently large to 
ignore boundary effects. Neumann boundary conditions are used and the equations are solved on a tetrahedral mesh. 
After having studied the convergence for different number of finite elements, we finally used a mesh with about 
$10^5$ elements. The exact number of elements differ depending on the size of the domain, determined by the value 
of $\omega$.

Initial data that is not too far away from the desired solution is needed to find a solution of the nonlinear equations. 
To overcome this difficulty we take advantage of the one-dimensional solutions and ``dress" them with a Gaussian profile 
in the transverse direction to obtain initial data good enough to find a first solution. The equations are then solved 
in sequence of increasing angular momentum, where each solution is provided as initial data for the next run. For each 
run the constant determining the wanted angular momentum is increased with a variable step with a minimal value of 0.001, 
until the full range $0 \leq \ell < 1$ is covered. A subset of 100 solutions of the total set of solutions is then 
saved.

The plots in Fig.\,4 show the density of the order parameter that is evaluated within the full, three-dimensional 
problem, for the same parameters as those of the intermediate plot of Fig.\,2, i.e., $n_0 U_0/\omega = 5/12$ ($\xi/R 
= 1/5$, $\xi/a_0 = 1.095$, $R/a_0 = \sqrt{30} \approx 5.477$), and for $\ell = 1/2$ and $3/4$. For $\ell = 1/2$ the 
density minimum is very low, which is a remnant of the ``dark" solitary wave of the one-dimensional/quasi-one-dimensional 
models. For $\ell = 3/4$, the density minimum is not as pronounced, again in agreement with the purely one-dimensional 
model, where the wave is ``grey", instead of ``dark", i.e., the density has a minimum which does not extend all the 
way to zero, though.

\begin{figure}
\includegraphics[width=8cm,height=8cm]{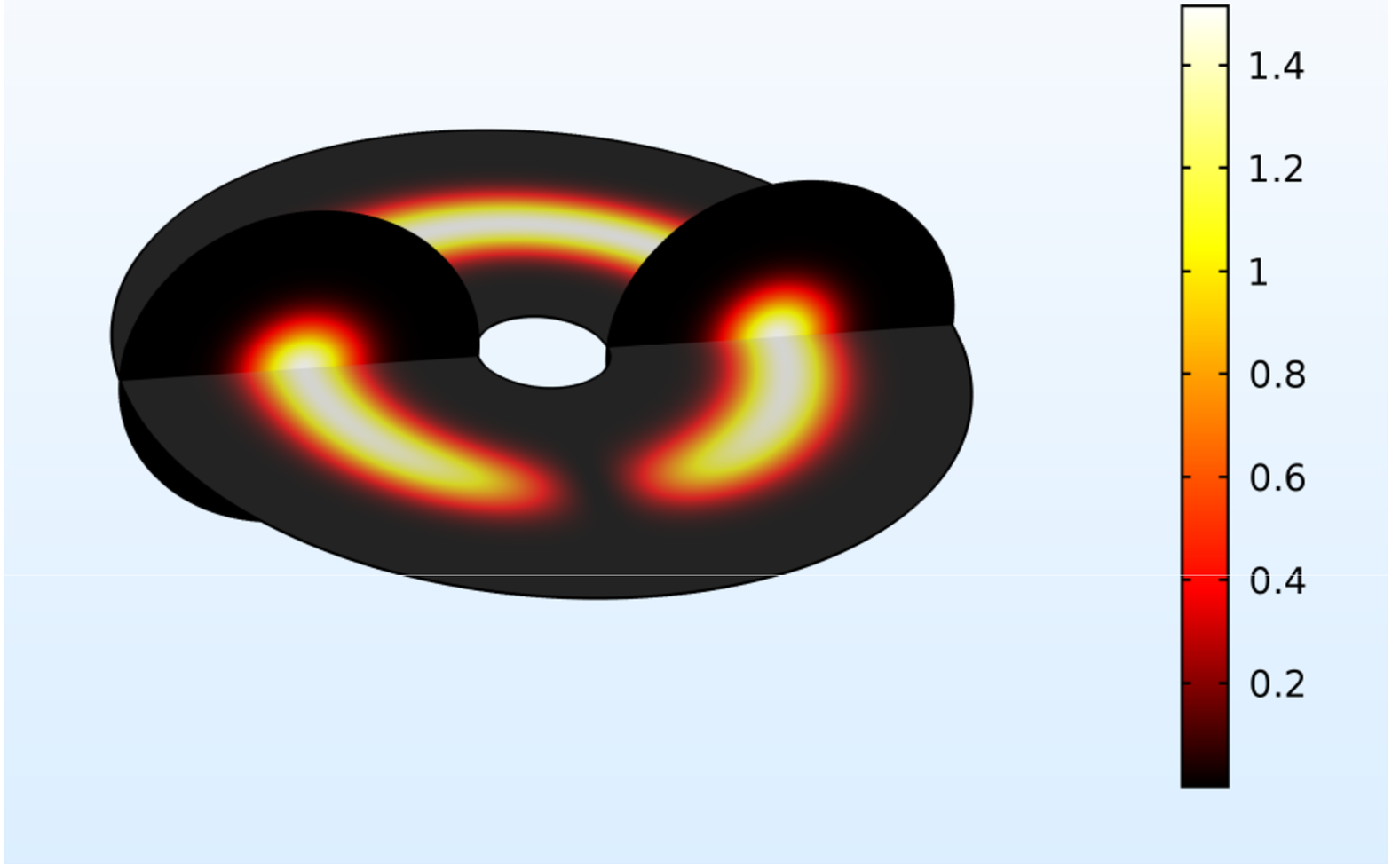}
\includegraphics[width=8cm,height=8cm]{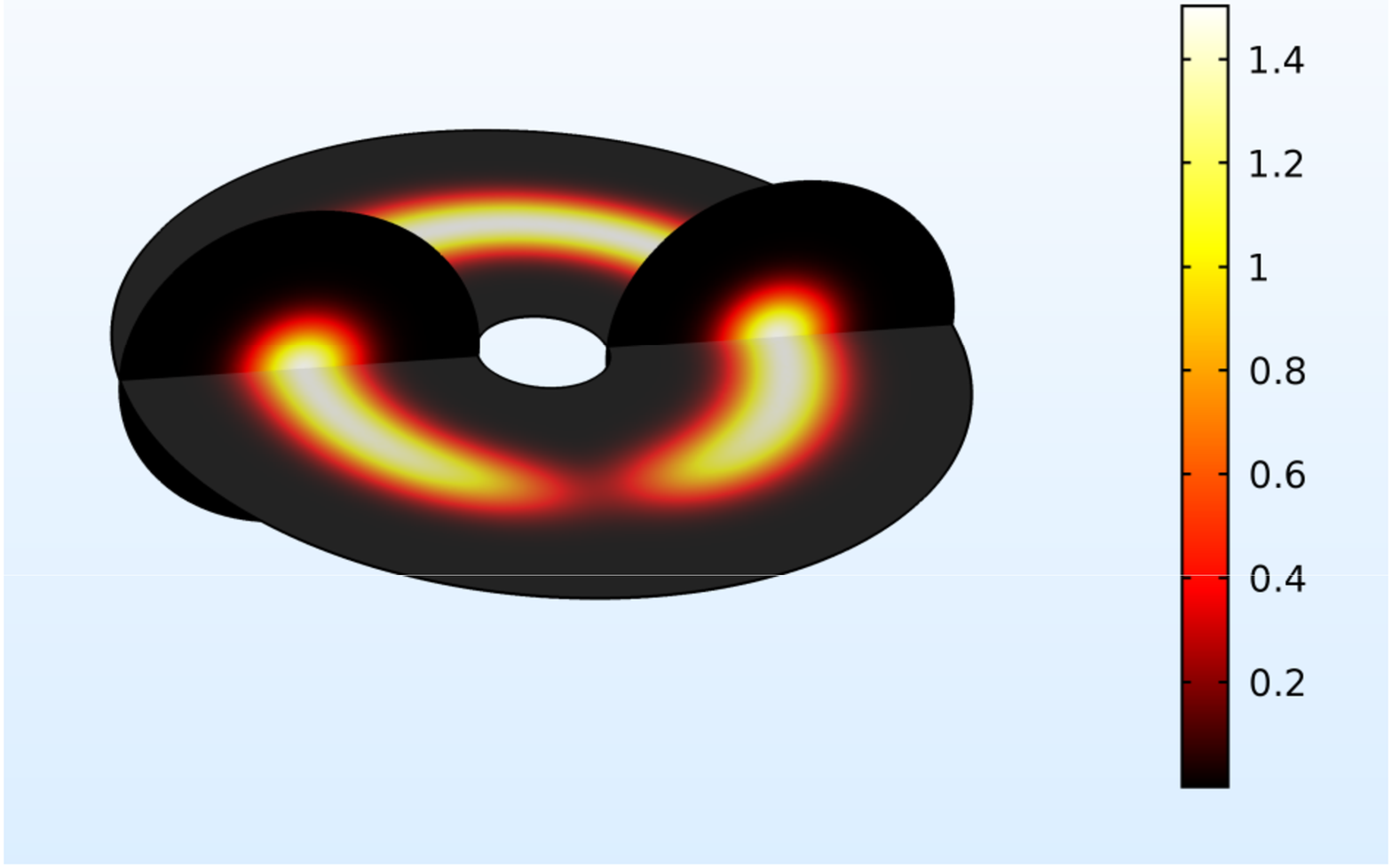}
\vskip1pc
\caption{(Colour online) The density $|\Phi|^2$ of the order parameter $\Phi$ of the full, three-dimensional model, 
on the $z=0$ plane, as well as along a perpendicular plane, for the same parameters as in the intermediate plot of
Fig.\,2, i.e., $n_0 U_0/\omega = 5/12$, $\xi/R = 1/5$, $\xi/a_0$ = 1.095 and $R/a_0 = \sqrt{30} \approx 5.477$. Here 
$\ell = 1/2$ (left) and $\ell = 3/4$ (right).}
\end{figure}

\section{Summary and conclusions}

The high degree of tunability of cold atomic systems allows us to build a superfluid system with properties which 
are designed at will. Tuning the trapping potential allows us to restrict the motion of the atoms either in two, 
or even in one spatial dimension and also to build topologically non-trivial traps. In the present study we have 
developed a model which is suitable in the case of a relatively tight confinement in two spatial dimensions, which 
results into quasi-one-dimensional motion. Apart from the assumption of relative tight confinement, it makes no
further assumptions, and it manages to reduce the three-dimensional problem into an effective one-dimensional, for 
any geometry, all the way between weak to strong (provided, of course that the mean-field approximation is still
valid) interatomic interactions. 

The model that we have derived is useful for practical purposes, since it reduces the computational effort that is
required in order to solve the corresponding nonlinear equation. Furthermore, our approach allows one to decouple 
and identify the three basic effects that enter this problem, namely the effect of the transverse confinement, of 
the curvature and of the interaction and see how each of them affects the system. Therefore it also provides insight 
into how all these various mechanisms affect the system, often in a competing way.

The quantitative comparison that we have made with the full numerical solution that we have found of the 
three-dimensional problem provides strong evidence that our model is rather accurate. While in the present study 
we have restricted ourselves to the case of the lowest mode in the transverse direction, one may generalize this 
method in order to include more modes. Such an approach would be even more accurate and it may be applicable in 
problems where the deviations from purely one-dimensional motion are substantial. As a result, it would be suitable 
in describing instabilities because of the deviations from purely one-dimensional motion see, e.g., Refs.\,\cite{Mur}.  

A whole new field, the so-called ``atomtronics", which focuses on the creation of atomic analogues to electronic  
devices has started to develop, see, e.g., \cite{GC}. These experiments are deeply in the Thomas-Fermi regime with 
respect to the transverse degrees of freedom of the condensate. Developing models like the one presented here is
certainly helpful, since they provide insight into these problems and also a relatively simple theoretical description. 
Another activity which is equally interesting is that of atomic waveguides (see, e.g., \cite{GO}), with obvious 
potential technological applications. Our model is applicable in such systems and it becomes especially interesting 
in the case where these waveguides bend, since it combines the effect of the curvature with that of the interaction. 

Finally, the present model is applicable in other quasi-one-dimensional systems. For example, considering the effect 
of the curvature combined with dipolar interatomic interactions \cite{dip} (instead of contact, considered here), may 
give rise to interesting effects. Equally interesting and important may also be the study of the combined effect of 
curvature with spin-orbit coupling \cite{soc}, which has been investigated in recent experiments.

\eject
{\it Acknowledgements.}
We acknowledge support from \"Orebro University, School of Science and Technology, and partly through RR 2015/2016.


\begin{thebibliography}{99}
\bibitem{integrals} To evaluate the integrals $B$ and $A$ respectively, we start with the integral 
$I(a) = \int_0^{2 \pi} {d \phi}/{(1 + a \sin \phi)^2}$, where $a = \kappa \rho$, and differentiate 
once and twice with respect to $a$.
\end{thebibliography}
\end{document}